\documentclass[prl,a4paper, amsfonts, amssymb, amsmath, reprint, showkeys, nofootinbib, twoside,superscriptaddress]{revtex4-2}

\usepackage[english]{babel}
\usepackage{graphicx}
\usepackage{subcaption}
\usepackage{amsmath}
\usepackage{amsfonts}
\usepackage{soul}
\usepackage[utf8]{inputenc}
\usepackage{xcolor}
\usepackage{physics}
\usepackage{overpic,color}
\usepackage{hyperref}
\hypersetup{
    colorlinks=true,
    linkcolor=blue,
    filecolor=magenta,      
    urlcolor=blue,
    citecolor = blue,
    pdftitle={Overleaf Example},
    pdfpagemode=FullScreen,
    }
\usepackage{dsfont}
\usepackage{mathtools}
\usepackage{bm}
\usepackage[normalem]{ulem}
\usepackage{lipsum}

\DeclareMathAlphabet{\pazocal}{OMS}{zplm}{m}{n}
\usepackage{isomath}

\newcommand{\eq}[1]{Eq.~(\ref{#1})}

\renewcommand{\BibitemShut}[1]{}

\usepackage[charter,expert]{mathdesign}

\begin{document}

\title{Diffraction patterns in attosecond photoionization time delay}

\author{Sajad Azizi}
\email[]{azizi.dajas@gmail.com}
\affiliation{Max Planck Institute for the Physics of Complex Systems, N\"{o}thnitzer Stra{\ss}e 38, D-01187, Dresden, Germany}
 \affiliation{Leibniz Supercomputing Centre, Boltzmannstra{\ss}e 1, 85748  Garching bei M{\"u}nchen, Germany}

\author{Mohamed El-Amine Madjet}
\affiliation{Department of Natural Sciences, Dean L. Hubbard Center for Innovation, Northwest Missouri State University, Maryville, Missouri 64468, USA}

\author{Zheng Li}
\affiliation{State Key Laboratory for Mesoscopic Physics and Collaborative Innovation Center of Quantum Matter, School of Physics, Peking University, Beijing 100871, China}
\affiliation{Collaborative Innovation Center of Extreme Optics, Shanxi University, Taiyuan, Shanxi 030006, China}
\affiliation{Peking University Yangtze Delta Institute of Optoelectronics, Nantong, Jiangsu 226010, China}

\author{Jan M. Rost}
\affiliation{Max Planck Institute for the Physics of Complex Systems, N\"{o}thnitzer Stra{\ss}e 38, D-01187, Dresden, Germany}

\author{Himadri S. Chakraborty}
\email[]{himadri@nwmissouri.edu}
\affiliation{Department of Natural Sciences, Dean L. Hubbard Center for Innovation, Northwest Missouri State University, Maryville, Missouri 64468, USA}
\date{\today}

\begin{abstract}
  Upon absorbing a photon, the ionized electron sails through the target force field in attoseconds to reach free space. This navigation probes details of the potential landscape that get imprinted into the phase of the ionization amplitude.  The Eisenbud-Wigner-Smith (EWS) time delay, the energy derivative of this phase, provides the navigation time relative to the time of the electron's ``free'' exit. This time is influenced by the diffraction of the electron from the potential landscape,  offering structural and dynamical information about interactions. If the potential has an intrinsic symmetry, a regular pattern in the time delay, including subpatterns of delays and advances, may occur from the diffraction process. The recent synthesis of a polyhedral fluorocarbon instigates the current study of photoionization from a cubic molecule. Our simulation of the EWS delay unravels rich diffraction motifs within $\pm$100 attoseconds in both energy and angular distributions. Averaging over the Euler angles from the laboratory to the molecular frame and over the photoelectron azimuthal direction indicates that the pattern should be discernible in ultrafast chronoscopy. The study benchmarks diffraction in molecular photoionization as a fundamental process which can be experimentally accessed through ultrafast time delay.
\end{abstract}\noindent

\maketitle

Novel technologies for the stability and control in the generation of pulses in the attosecond extreme ultraviolet (XUV) and improvements in data acquisition techniques to better statistics allow the tracking and maneuvering of electron dynamics in matter~\cite{hui2024,tang2024}. This continues to trailblaze the ultrafast access to photoionization (PI) dynamics~\cite{calegari2024,calegari2016}. In two-photon interferometric tracks, namely, reconstruction of attosecond beating by interference of two-photon transition (RABBITT)~\cite{paul2001,kluender2011} and streaking~\cite{keinberger2004,seiffert2017}, XUV pump pulses are synchronized to a controlled delay with an infrared (IR) probe pulse. Pulses from free-electron laser sources have recently been employed to access soft x-ray energies~\cite{driver2024} as well. By measuring the attosecond time delay in PI, studies accessed the anisotropy in molecular potentials~\cite{ahmadi2022} and the size-resolved effect of the clusters, establishing an approach to follow exciton localization~\cite{gong2021}. Exploring the dynamics of molecular shape resonances~\cite{heck2021} and interference between centers in diatomic molecules~\cite{heck2022} is performed by accessing PI delays in electron-ion coincidence techniques.

The basic quantum amplitude $\bra{\boldsymbol{f}}\hat{O}\ket{\boldsymbol{i}}$ of interaction, represented by an operator $\hat{O}$, of a charged particle with a molecular target garners key structural and spectroscopic information of the target. The interaction causes a transition of the particle from an initial $\ket{\boldsymbol{i}}$ to a final $\ket{\boldsymbol{f}}$ state, where the continuum-to-continuum and bound-to-continuum (when the particle is electron) transitions are respectively scattering and ionization. This amplitude is a complex quantity that has a modulus, $|\bra{\boldsymbol{f}}\hat{O}\ket{\boldsymbol{i}}|$, and a phase, $\Phi$ = arg($\bra{\boldsymbol{f}}\hat{O}\ket{\boldsymbol{i}}$), both being fundamental quantities. Routinely observable is the squared-modulus of the amplitude, proportional to the number (intensity) of the particles scattered or ionized which maps to the cross section. On the other hand, the energy derivative $\partial\Phi/\partial E$ of the phase provides a Hermitian observable, the Eisenbud-Wigner-Smith (EWS) time delay induced through the interaction~\cite{Eisenbud1948,Wigner_1955,smith1960}. The latter can be understood as the delay incurred while scattering off or ionizing from the target's electrostatic potential.

For systems of certain geometric or structural symmetry, these observables can engender diffraction motifs. The effect originates from systematic overlaps, induced by the symmetry, of scattered/ionized spherical waves with patterns of path differences. Studies of diffraction in electron scattering cross sections are abundant involving molecules~\cite{saha2022,amini2020}, nanostructures~\cite{ponce2021}, surfaces~\cite{Kienzle2012}, and crystals~\cite{gemmi2019}. However, diffraction in photoelectron intensity was only accessed for a spherical C$_{60}$ molecule~\cite{ruedel2002} and was predicted for the atomic endohedral C$_{60}$~\cite{mccune2009}. The diffraction in EWS delay of PI is studied only for C$_{60}$~\cite{magrakvelidze2015}. This leaves open an almost entire field of inquiry on {\em diffraction in time-domain} within the PI landscape.

In PI, the photoelectron is produced inside the target, propagates through the target potential on its way to the detector. With the increase of photon energy, hence the photoelectron momentum, the de Broglie wavelength of the photoelectron shortens. For a spherical target, whenever integer multiples of photoelectron waves fit the target diameter, maximum cancellations in the overlap integral of the amplitude occur. These interference conditions result in a series of minima (dark spots) in the amplitude. However, for PI of a spherical system, the diffraction as a function of the emission angle is forbidden~\cite{AngDistPI}, since the path length through the target potential is identical in all directions. Incidentally, this is not true for electron scattering from a spherical target~\cite{aiswarya2024}, since a preferential direction of incidence breaks the symmetry. But, for a target with non-spherical symmetry, diffraction in both photoelectron energy and emission angle is expected to occur. 
\begin{figure}[ht]
    \includegraphics[width=\columnwidth]{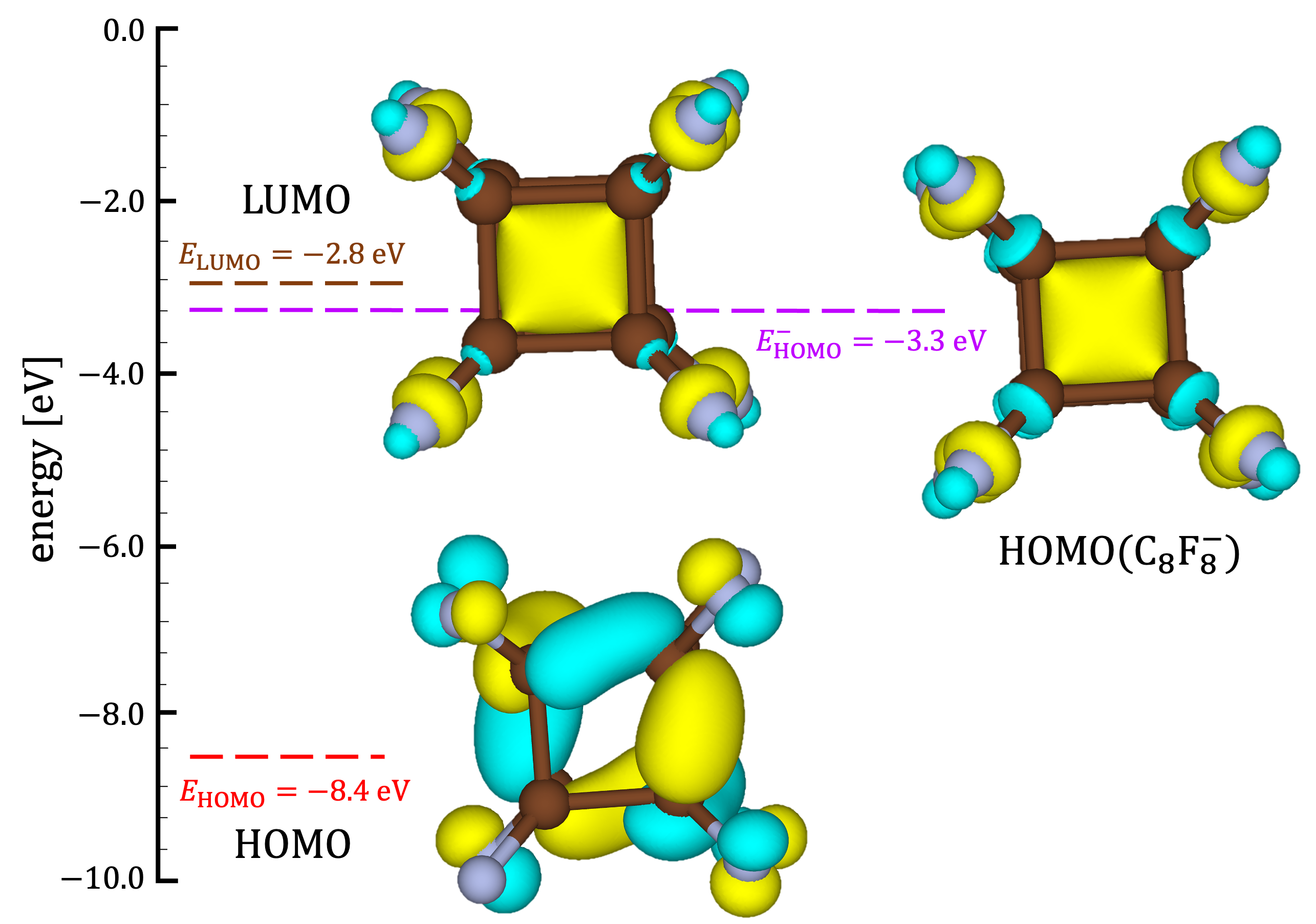}
    \captionsetup{justification=raggedright,singlelinecheck=false}
    \caption{Contour plots for HOMO and LUMO of C$_8$F$_8$ and HOMO of C$_8$F$_8^-$ computed at the B3LYP/6-311++G(d,p) level of theory and shown with energies.}
    \label{figs:fig1}
\end{figure}
\begin{figure*}[ht]
    \includegraphics[width=0.8\textwidth]{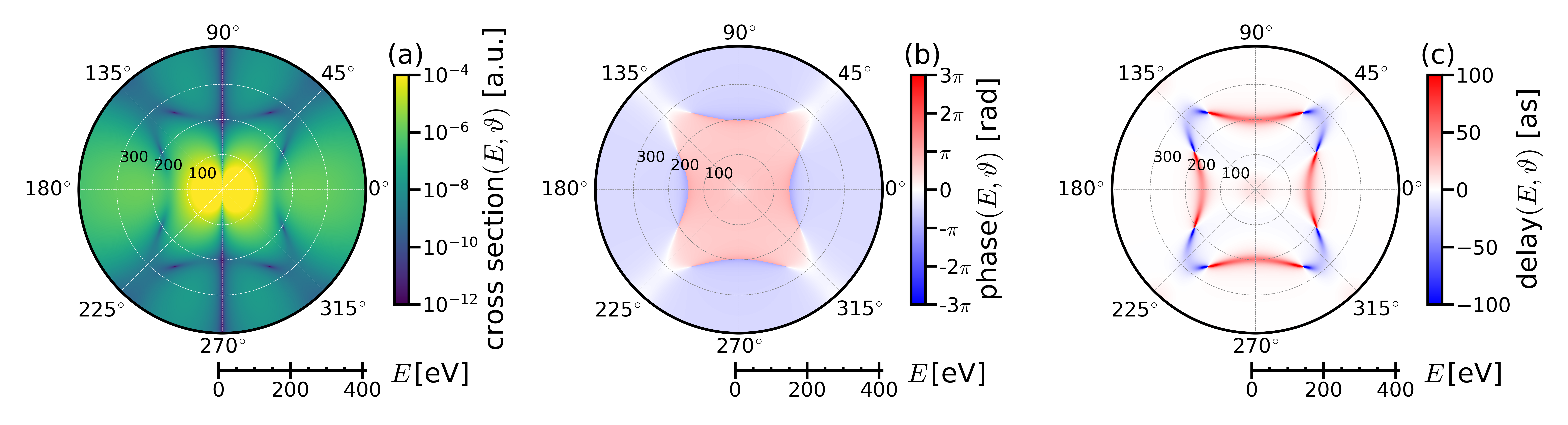}
    \captionsetup{justification=raggedright,singlelinecheck=false}
    \caption{Polar angular distribution images, with photoelectron kinetic energy plotted in the radial direction, of cross section (a), phase (b) and time delay (c) for $\beta = 0$ and $\varphi = 0$.}
    \label{figs:fig2}
\end{figure*}

The current study ratifies this expectation. Our prediction of multidimensional diffraction should motivate experiments to capture the resulting patterns. While the photoelectron intensity can be accessed with synchrotron lights~\cite{gerlach2023,karaev2024}, the more contemporary ultrafast measurements~\cite{heck2021,ahmadi2022} by RABBITT or streaking techniques can be applied to observe diffractions in time delay. 

A perfect target system to this purpose is found in the recently synthesized, and confirmed by x-ray crystallography, polyhedral perflurocubane C$_8$F$_8$ molecule, whose cubic structure is held by eight C atoms located at the corners of the cube~\cite{Sugiyama_2022}. The molecule can efficiently accept an electron in its LUMO, derived from its eight C–F $\sigma^\ast$ orbitals, to form a radical anion C$_8$F$_8^-$. The delocalized sequestration of this electron in the cube was probed by matrix-isolation electron spin resonance (ESR) spectroscopy. Following Ref.\,\cite{Sugiyama_2022}, we have also performed quantum chemical simulations of the molecular structure based on density functional theory of the level of B3LYP/6-311++G(d,p) for neutral C$_8$F$_8$ in the solvent tetrahydrofuran. We used optimization and SCF by employing Gaussian~\cite{frisch2019} with D3 dispersion correction. We extended the calculation to anionic C$_8$F$_8^-$. The corresponding HOMO and LUMO orbitals of the neutral molecule and the HOMO orbital of the anion are presented in Figure~\ref{figs:fig1}. Simulations confirm that the LUMO of the neutral and HOMO of the anion predominantly delocalize in the cubical region. Thus, PI from the anion's HOMO level, or from an excited level of the neutral system, will experience cubic symmetry of a diffractor. This should produce angle and energy domain diffraction signatures in both the intensity and EWS delay signals of PI.

We focus on this delocalized cubic orbital of perflurocubane in the single-active-electron approximation. 
A potential with cubic symmetry is adopted to model the molecular system; see Supplementary Material (SM)~\cite{supplemental}. Given the parameters chosen, the system manifests a ground state energy $E_0 = -2.81$\,eV, well matching the energy $-2.8$\,eV \cite{Sugiyama_2022} of the LUMO of C$_8$F$_8$ (Fig.\,\ref{figs:fig1}). Obviously, the model can also be readily adapted for the anionic HOMO state, with a slightly lower energy (Fig.\,\ref{figs:fig1}), that will trivially shift the spectrum.
The laser-molecule coupling Hamiltonian in length gauge reads
\begin{equation}
 \label{eq:hamil}
 H = \boldsymbol p^2/2 + V(\boldsymbol r) +\boldsymbol r \cdot \boldsymbol F,
\end{equation}
where $\boldsymbol F$ denotes the electric field of the linearly polarized laser and the dipole transition operator $\hat{\mu} = \boldsymbol r \cdot \boldsymbol F$; we assumed $|\boldsymbol F|=1$. Details of the theory and computation are given in SM~\cite{supplemental}. We provide a succinct account below.

The dipole matrix elements in the PI amplitude can be expressed in the length gauge as,
\begin{equation}\label{eq:dipole}
    \mathcal{D}^{(-)}(\boldsymbol k) = \left \langle \Psi^{(-)}_{f}(\boldsymbol k,\boldsymbol r) | \boldsymbol r \cdot \boldsymbol F| \Psi_{i}(\boldsymbol r) \right \rangle,
\end{equation}
where $\Psi_{i}$ represents the initial state and $\Psi^{(-)}_{f}$ the final state wavefunction. In Eq.\,\eqref{eq:dipole}, $\hat{\boldsymbol k} \equiv (\vartheta, \varphi )$ denotes the direction of photoelectron emission at the polar and azimuthal angle, respectively, that define the molecular frame (MF) (Fig.\,S1 in SM \cite{supplemental}). The magnitude $k$ of the photoelectron momentum relates to its kinetic energy $E$ as $k = \sqrt{2E}$ in a.u.  The squared modulus of the matrix elements accounts for the cross section. The EWS delay, $\tau(\boldsymbol k)$, is computed from the energy derivative of the phase of the matrix elements. 

Our main results are presented in MF to make them amenable for experiments. Fig.\,S1 in SM \cite{supplemental} also displays the laboratory frame (LF) defined by the photon polarization direction and the Euler angles ($\alpha$, $\beta$ and $\gamma$), that transform from LF to MF. Since we chose linearly polarized light, $\gamma$ automatically vanishes and is therefore ignored. $\alpha$ and the polarization angle $\beta$, however, remain sensitive.

In order to capture the evolution of the diffraction pattern in PI from a spherical to a cubic potential, we carry out a numerical experiment. We introduce a degree-of-squareness parameter $s$ following Ref.\,\cite{Guasti_1993}; see Fig.\,S3 in SM~\cite{supplemental} for details. The shape of the potential evolves via $s$ from a sphere ($s=0$) to a cube ($s=1$). As shown, corresponding results of both cross section and EWS delay evolve dramatically. Spherical fringes for $s=0$ progressively deform through richer patterns as the shape approaches $s=1$. This demonstrates the strong sensitivity of the photoelectron diffraction to the symmetry. Results presented below pertain to the cubic ($s=1$) case.

Figure~\ref{figs:fig2} presents images of the photoelectron polar distribution, with $E$ plotted radially, for cross section, phase and time delay. We choose both $\beta$ and $\varphi$ to be zero, which makes $\vartheta = 0,\pi$ to coincide with the polarization axis and $\alpha=\gamma=0$. Note that the cross section, panel (a), peaks immediately above the ionization threshold ($E=0$). This region is strongly influenced by the shape resonance from the electron's confinement through the repulsive centrifugal barrier potential created by higher angular momentum components of the continuum wave. This region is found to be surrounded at higher energies by an astroid (mathematically, a hypocycloid with four cusps) shaped minimum profile from diffraction. The astroid is found slightly stretched in the laser propagation axis ($\vartheta = \pi/2, 3\pi/2$). We note a minimum in the cross section along this axis. This is due to the fact that the dipole coupling with linearly polarized light contains $\cos(\vartheta)$ which renders $\mathcal{D}$ zero for $\vartheta =\pi/2, 3\pi/2$ -- an effect to which the phase and time delay images, (b) and (c), are insensitive (see SM~\cite{supplemental} for details).   

Fig.\,\ref{figs:fig2}(c) for time delay, the energy gradient of the phase in (b), mimics~\cite{ji2024} the cross section image in (c) for the astroid profile of diffraction. Particularly, it reveals a positive delay and a negative delay (time advance) substructure symmetry within this profile. In general, delays are seen around $\vartheta = n(\pi/2)$ polar emissions while advances appear around $\vartheta = (2n+1)(\pi/4)$ emissions, where $n$ are positive integers including zero. To further assess this effect we express the complex amplitude $\mathcal{D}$ as $(R + i I)$ with $R$ and $I$ being the real and imaginary components. From Eqs.\,(S6) and (S7) of SM~\cite{supplemental} we thus obtain the time delay as,
\begin{equation}\label{CStoTD}
  \tau(\boldsymbol k) \sim \frac{RI' - R'I}{\sigma},
\end{equation}
where the energy derivatives are $R'$ and $I'$ for brevity. 
\begin{figure}[bp]
    \centering
    \includegraphics[width=\columnwidth]{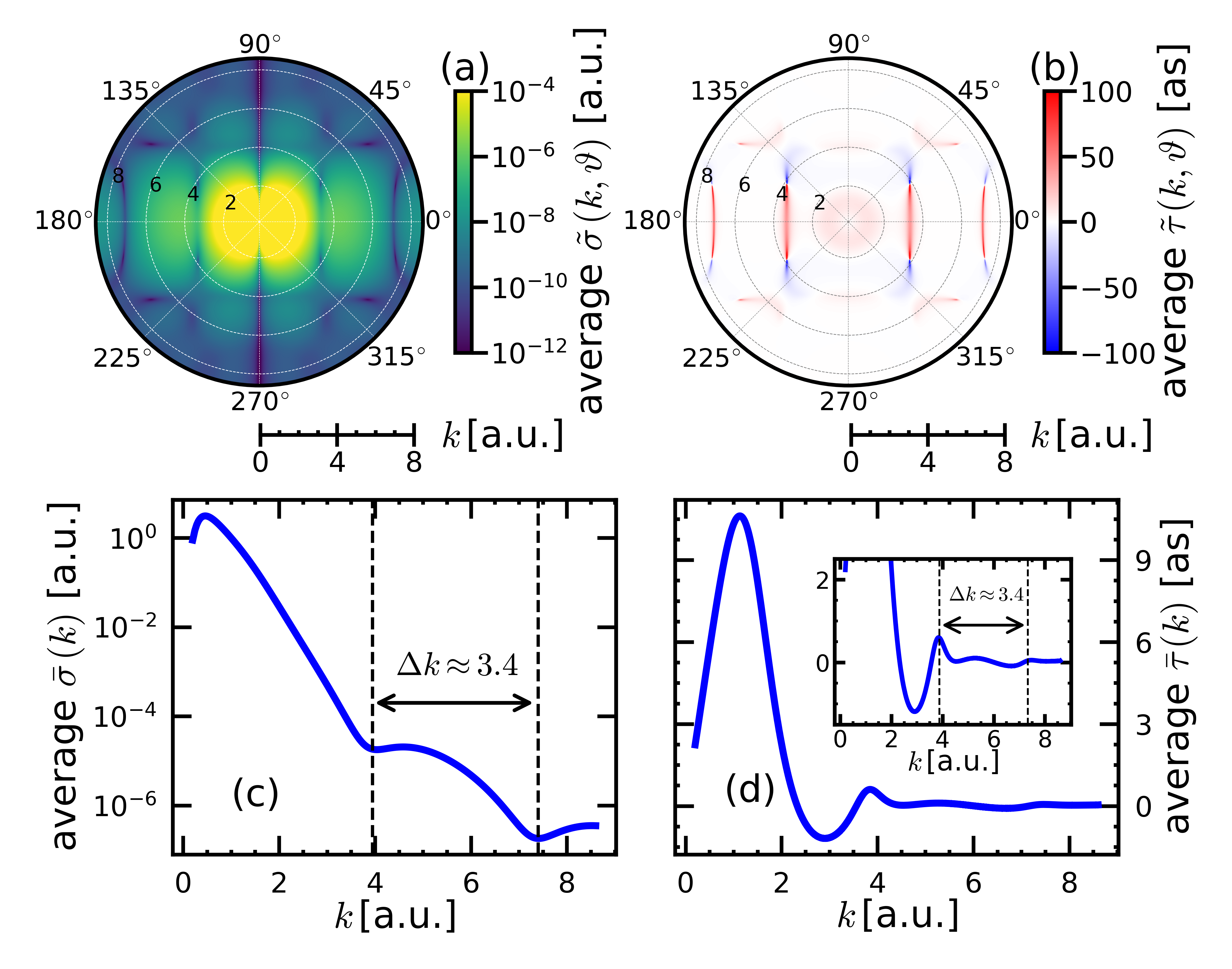}
    \captionsetup{justification=raggedright,singlelinecheck=false}
    \caption{Extended energy polar angular distribution images averaged in $\varphi$ for cross section (a) and time delay (b) for $\beta = 0$. (c) and (d) are obtained after a further $\vartheta$ averaging; inset in (d) for zoom-in. All results are plotted in $k$ where the fringe separation, marked in dashed lines, $\Delta k$ = 3.4 a.u.\ yields $2 \pi/\Delta k = 1.85$ a.u.}
    \label{figs:fig3}
\end{figure}
\begin{figure*}[!htbp]
    \centering
    \includegraphics[width=0.8\textwidth]{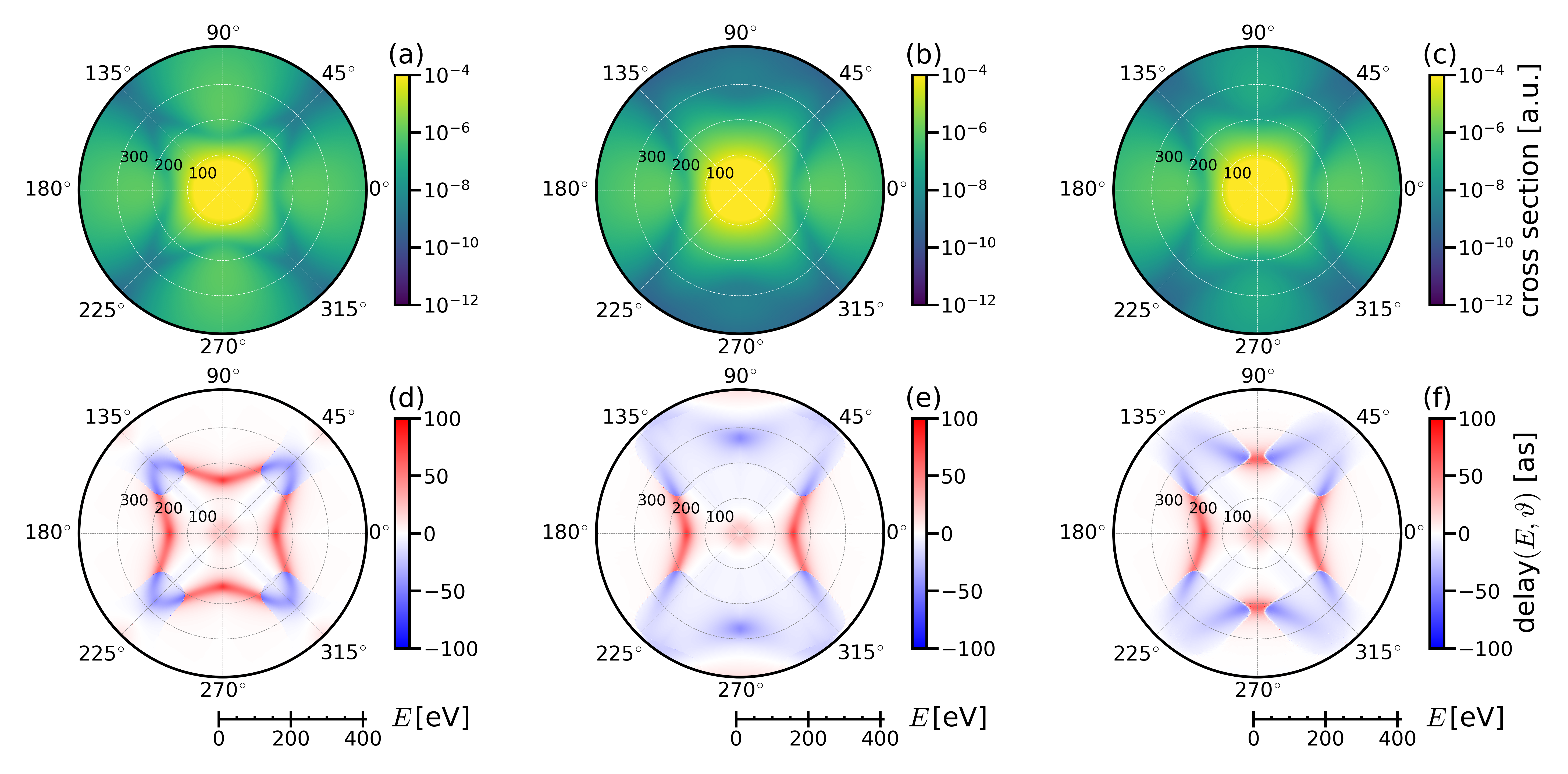}
    \captionsetup{justification=raggedright,singlelinecheck=false}
    \caption{Cross section and time delay polar images for different choice of $\varphi$ direction but averaged on Euler angles $\alpha$ and $\beta$. We chose $\varphi = 0,\pi/2$ (a,d), $\varphi=\pi/4$ (b,e), and $\varphi=\pi/3,\pi/6$ (c,f)}
    \label{figs:fig4}
\end{figure*}

Equation~\eqref{CStoTD} clearly indicates that a minimum in cross section will translate to an extremum in the time profile. But, whether this extremum is a delay or an advance will depend on the sign of the numerator in \eq{CStoTD}. A careful inspection at the astroid profile in the cross section, Fig.\,2(a), reveals that the profile has shallow and deep minima, a substructure which corresponds to the identical angular symmetry in the time delay. In other words, shallow minima are seen around $\vartheta = n(\pi/2)$, while deep minima are found around $\vartheta = (2n+1)(\pi/4)$. This implies that at the deep minima regions of a fringe in cross section the numerator of \eq{CStoTD} becomes negative to produce time-advanced emissions.

We may understand this phenomenon by motivating a physical picture. Within a diffraction fringe profile of the cross section, the shallow minima are bright spots, while the deep minima are dark spots. Since, as noted earlier, dark spots are produced when integer multiples of photoelectron waves fit the diffractor size along the emission direction, these can be thought of ``resonance'' states in the continuum. This is equivalent to the Cooper minimum where a similar ``diffraction'' criterion is met, but by the bound-state radial wave with nodes~\cite{Cooper_1962}. Obviously, when such a resonance condition is attained, the photoelectrons rapidly transition to the continuum to occupy the resonance state, causing a time-advanced exit. Conversely, for bright spots (shallow minima) in the cross section, the opposite happens, that is, the photoelectron results in a delayed emission there. We may also note an above-threshold positive delay in Fig.\,\ref{figs:fig2}(c) due to transient capture of the electron within the centrifugal barrier~\cite{gongPRX2022}.

In Figure~\ref{figs:fig3} we extend the energy to cover approximately the lower third of the soft x-ray domain, while setting $\beta = 0$. The cross section (a) and time delay (b), computed after averaging over $\varphi$, reveal richer diffraction fringes, which are commensurable between these images. Note that here we plot the images in $k$ which morphs the fringe shapes to be somewhat straightened out from the cusps in $E$. We then add the $\vartheta$ averaging to express the signals as a function of only $E$ in panel (c) and (d). These curves prominently show the above-threshold resonance from the centrifugal barrier (c) and the resulting delay (d). Remarkably, each curve yields two discernible extrema (fringes) corresponding to a momentum separation ($\Delta k$) of 3.4 a.u. As expected from a diffraction mechanism in the photoelectron momentum, the Fourier reciprocal $2\pi/\Delta k$ = 1.85 a.u., being the angular averaged size of the diffractor, closely approaches $L = 1.7$ a.u.\ of the cubic potential adopted. The weak appearance of the fringes in (c) and (d) are due to cancellations in averaging over the details of angular subpatterns.  

Images in Figure~\ref{figs:fig4} are simulated for a selections of $\varphi$ while averaging the Euler angles. These are displayed on a limited energy range as in Fig.\,\ref{figs:fig2} to capture the $\varphi$-dependence of the astroid pattern. Since averaged over the Euler angles, these results display the polar angular distributions of photoelectron cross section and time delay in MF. Results also exhibit an azimuthal symmetry relationship $\varphi \rightarrow (\pi/2 - \varphi)$ of a period of $\pi/2$ as expected of a cubic symmetry. Note that the astroid for $\varphi = 0,\pi/2$ emerges as completely symmetric in shape and intensity. However, the top and the bottom cusps are found to evolve through a cycle of gradual intensity-fading and energy blue-shift and then brightening and red-shift as $\varphi$ goes from zero to $\pi/2$. Since this polar direction $\vartheta=\pi/2, 3\pi/2$ lies on the azimuthal plane (Fig.\,S1), the effect evolves within a period of $\pi/2$ of $\varphi$ from the cubic symmetry. In contrast, the left and right cusps oriented along $\vartheta=0,\pi$ being orthogonal to the azimuthal plane remain roughly unchanged as $\varphi$ alters.
\begin{figure}[ht]
    \centering
    \includegraphics[width=\columnwidth]{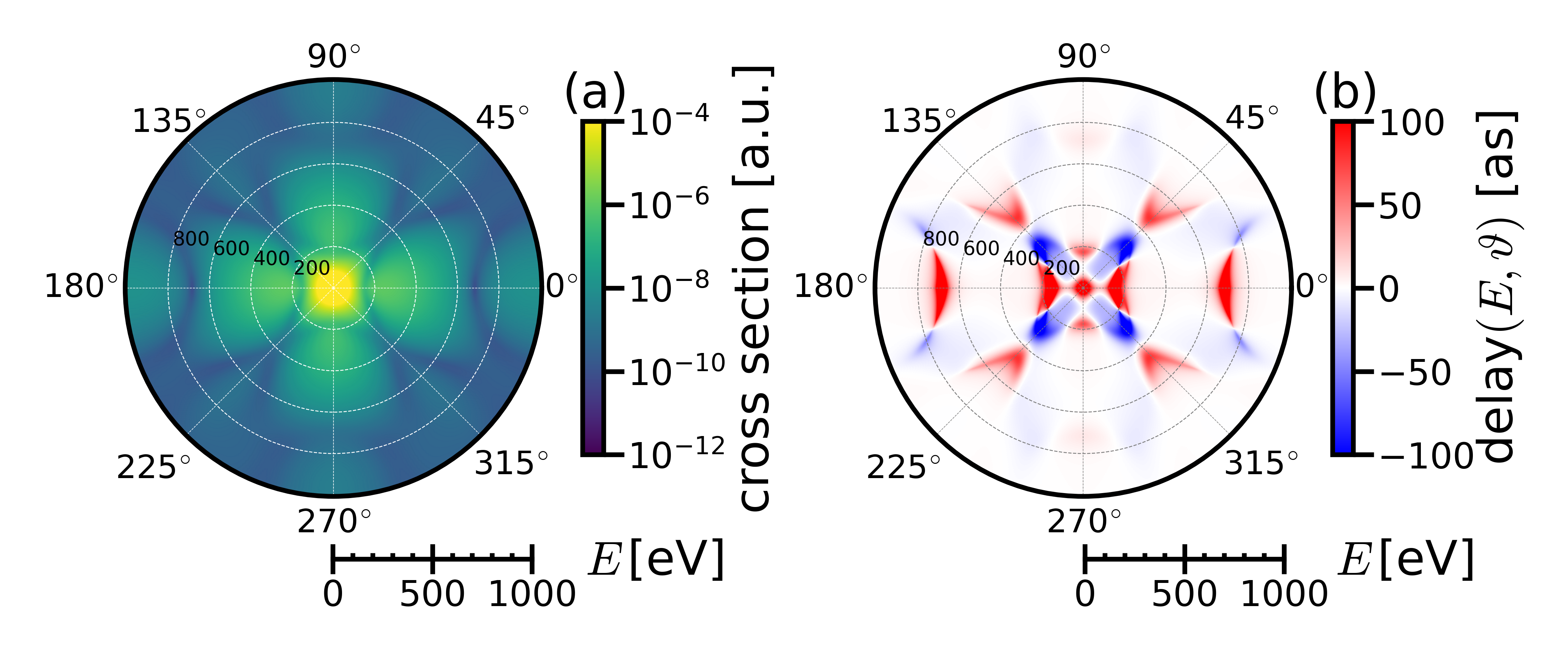}
    \captionsetup{justification=raggedright,singlelinecheck=false}
    \caption{Diffractograms of cross section (a) and time delay (b) in the molecular frame obtained by averaging both on Euler angles and the azimuthal direction.}
    \label{figs:fig5}
\end{figure}

Figure~\ref{figs:fig5} unveils the results, after averaging the azimuthal directions, of the angular distributions solely on the polar plane for cross section in panel (a) and time delay in panel (b). The energy scale extends to 1 keV. These 2D diffractograms in MF are independent of the choice of an azimuthal direction and therefore, can be accessible in the experiment. The cross section's diffraction patterns exhibit systematic deeper and shallower minima. The time delay image reciprocates this pattern, while remarkably, the maxima of the delay and advance show consistent increase, ranging closely between $\pm$100 as. This is plausibly a consequence of the azimuthal averaging. This boost in the magnitude of the temporal diffraction and a richer structure even below 400 eV may facilitate ultrafast measurements. To estimate the additional delay introduced by the probe pulse in a streaking measurement, the EWS result can be augmented by, for instance, a classical trajectory Monte Carlo propagation approach~\cite{biswas2021}, for higher accuracy. One may also test fundamental predictions~\cite{saalmaan2021}.

Ref.\,\cite{Sugiyama_2022} finds that the observed radical anion C$_8$F$_8^-$ rotates rapidly in the glass matrix at 77 K. For a free-oriented molecule, measurements will automatically incorporate angular averaging. We remark that results presented in Figs.\,\ref{figs:fig4} and \ref{figs:fig5} are likely candidates for photoelectron intensity and delay measurements using, respectively, synchrotron light sources or ultrafast pump-probe laser pulses on the track of RABBITT or streaking. In order to enhance the signal of the diffraction, one may eventually need molecules ordered on the matrix with lateral dimensions in microns.  If the molecules do not order, one may be able to look at the extended edge x-ray absorption signal and black out or minimize the nearest-neighbor interactions. If technology may support the anions in vapor phase, it may be possible to measure photo-intensity and delay signals in standard anion detachment technique~\cite{DeVine2018,Bragg2004,Bragg2003}

To conclude, we have carried out a detailed computation and analysis of the photoelectron energy and angular distribution of the outer level of a polyhedral fluorocarbon molecular anion, known to be very stable among other cubane anions~\cite{martins2023}. Resulting cross sections and time delays exhibit profound diffraction patterns from the molecule's symmetry. This diffraction is a fundamental effect which should be generally found in PI of molecules with stable symmetries~\cite{ghosh2023}. Structures in cross sections can be accessed by photoelectron intensity measurements with synchrotron lights. More attractively, our results pioneer the prediction of energy and angle differential diffraction phenomenon in the PI time delay. It was accomplished on the EWS track, which has shown success to corroborate with measurements~\cite{magrakvelidzePR2015, biswas2021, mukherjee2024}. Thus, the predicted temporal structure of diffraction is worth pursuing in the ultrafast laser pump-probe technique.

Z.L.\ acknowledges funding from National Natural Science Foundation of China (Grants No.~12174009, 12234002, 92250303), the National Key R\&D Program (Grant No.~2023YFA1406801) and Beijing Natural Science Foundation (Grant No.~Z220008). H.S.C.\ acknowledges support from the US National Science Foundation (Grant Nos.~PHY-2110318 and CNS-1624416).



\end{document}